\documentclass[lettersize,journal]{IEEEtran}
\usepackage{amsmath,amsfonts}
\usepackage{algorithmic}
\usepackage{algorithm}
\usepackage{array}
\usepackage[caption=false,font=normalsize,labelfont=sf,textfont=sf]{subfig}
\usepackage{textcomp}
\usepackage{stfloats}
\usepackage{url}
\usepackage{verbatim}
\usepackage{graphicx}
\usepackage{cite}
\begin{document}

\title{A Deep Unfolding Framework for Diffractive Snapshot Spectral Imaging}

\author{Zhengyue Zhuge, Jiahui Xu, Shiqi Chen, Hao Xu, Yueting Chen, Zhihai Xu, Huajun Feng 
\thanks{The authors are with the State Key Laboratory of Extreme Photonics and Instrumentation, Zhejiang University, Hangzhou 310000, China. E-mail: \{zgzy, xu\_jiahui, chenshiqi, xuhao\_optics, chenyt, xuzh, fenghj\}@zju.edu.cn}
}



\maketitle

\begin{abstract}
Snapshot hyperspectral imaging systems acquire spectral data cubes through compressed sensing. Recently, diffractive snapshot spectral imaging (DSSI) methods have attracted significant attention. While various optical designs and improvements continue to emerge, research on reconstruction algorithms remains limited. Although numerous networks and deep unfolding methods have been applied on similar tasks, they are not fully compatible with DSSI systems because of their distinct optical encoding mechanism. 
In this paper, we propose an efficient deep unfolding framework for diffractive systems, termed diffractive deep unfolding (DDU). 
Specifically, we derive an analytical solution for the data fidelity term in DSSI, ensuring both the efficiency and the effectiveness during the iterative reconstruction process. 
Given the severely ill-posed nature of the problem, we employ a network-based initialization strategy rather than non-learning-based methods or linear layers, leading to enhanced stability and performance.
Our framework demonstrates strong compatibility with existing state-of-the-art (SOTA) models, which effectively address the initialization and prior subproblem. 
Extensive experiments validate the superiority of the proposed DDU framework, showcasing improved performance while maintaining comparable parameter counts and computational complexity. These results suggest that DDU provides a solid foundation for future unfolding-based methods in DSSI.
\end{abstract}


\section{Introduction}
\IEEEPARstart{T}{he} spectral characteristics of scenes have garnered widespread attention across various fields, including agriculture \cite{wang2021combining}, environmental monitoring \cite{adao2017hyperspectral}, object tracking \cite{xiong2020material}, and face recognition \cite{zhang2021hyperspectral}. Traditional hyperspectral imaging systems require temporal or spatial scanning to capture spectral images, making the process time-consuming and dependent on bulky hardware. This limitation restricts their applicability in dynamic scenes and compact devices (e.g., surveillance cameras and smartphones). 

Recently, several snapshot imaging systems have been developed for fast and compact spectral imaging. These systems use optical components to encode 3D spectral cubes into 2D images, which are then recovered into hyperspectral images (HSIs) through reconstruction algorithms. 

A conventional RGB camera can be considered as a snapshot spectral imaging system, as filter arrays placed in front of the sensor trade spatial resolution for spectral information.
Competitions like New Trends in Image Restoration and Enhancement (NTIRE) challenges \cite{arad2018ntire, arad2020ntire, arad2022ntire} have driven the development of numerous network-based methods to recover hyperspectral cubes from RGB images. However, estimating HSIs from clear RGB images often encounters issues of metamerism \cite{fu2024limitations}.

\begin{figure}[t]
  \centering
  \includegraphics[width=1.0\linewidth]{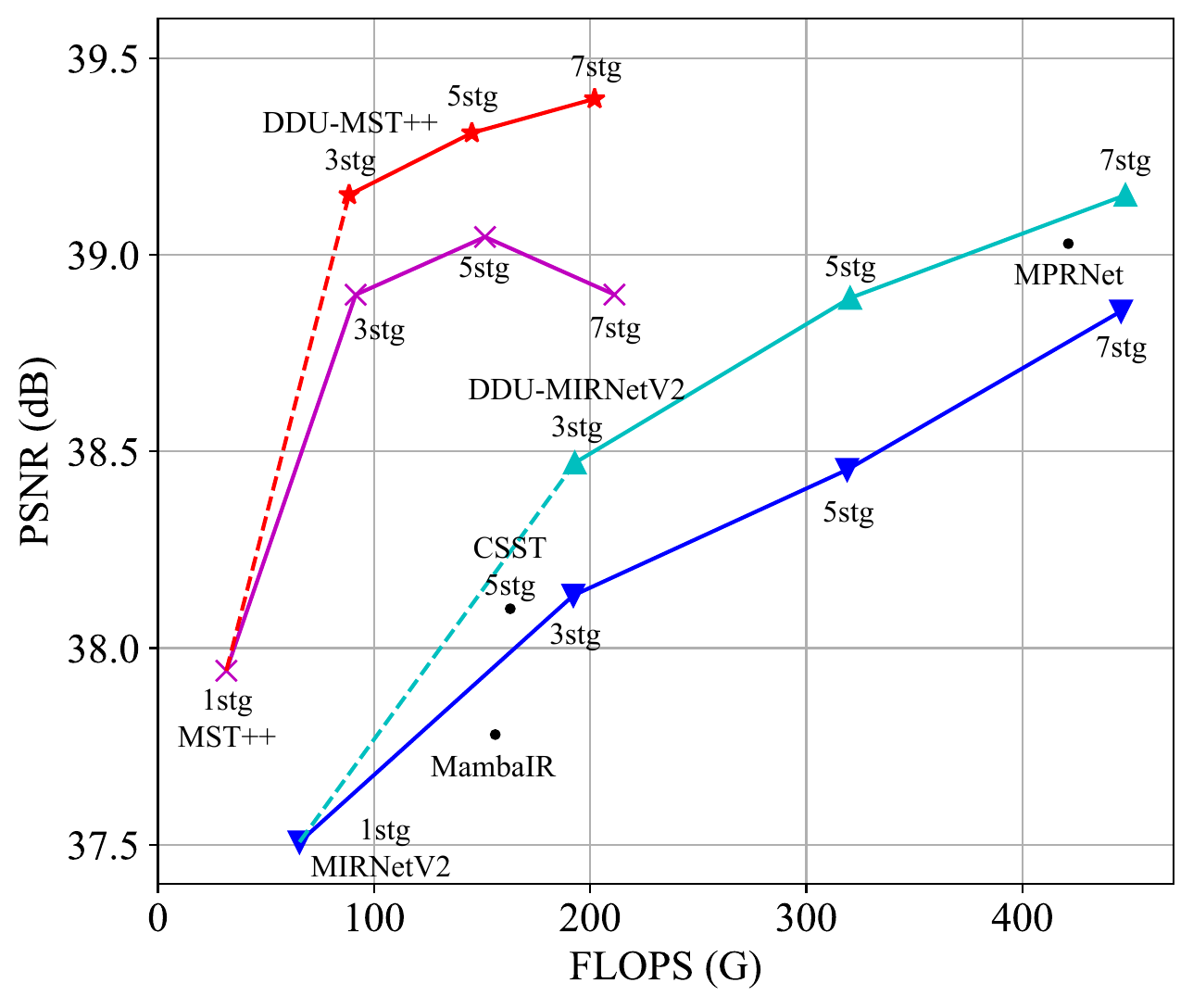}
  \caption{PSNR vs. FLOPS comparison on a mixed test set comprising samples from the ICVL \cite{arad2016sparse} and VNIRSR \cite{xu2025two} datasets. For MST++ and MIRNetV2, ``stg" denotes the number of SST and RRG blocks, respectively. For deep unfolding methods, ``stg" represents the total number of initialization networks and iterative blocks.}
  \label{fig: Scatter}
\end{figure}

Coded aperture snapshot spectral imaging (CASSI) systems \cite{gehm2007single} encode input spectra using a binary mask and then disperse them onto the sensor. Many improvements have been proposed on optical design \cite{wu2011development, parada2017colored} and reconstruction algorithms \cite{cai2022mask, wang2022snapshot}. However, CASSI has a relatively long system length because of multiple relay lenses and suffers from low light efficiency due to mask occlusion. These limitations hinder its practical applications. 

Diffractive snapshot spectral imaging (DSSI) systems address these challenges by engineering the point spread functions (PSFs) of the imaging system, typically using diffractive optical elements (DOEs).
The diffracted rotation system, proposed by Jeon et al. \cite{jeon2019compact}, designs wavelength-dependent PSFs using a single DOE. Subsequent research \cite{xu2024video} integrated diffracted rotation systems with rear lenses to improve imaging quality and shorten the system length. Compared to CASSI, DSSI systems are significantly more compact and lightweight. Moreover, because of the absence of mask occlusion, they introduce virtually no additional optical energy loss and consequently exhibit lower noise levels. These advantages make DSSI particularly well-suited for deployment on mobile platforms and in environments with constraints on space and illumination.

Although DSSI systems have shown promising optical characteristics, research on reconstruction algorithms specifically designed for DSSI remains limited. Recent studies of CASSI have incorporated optical masks as network inputs \cite{cai2022mask} and have explored a lot in deep unfolding methods \cite{wang2020dnu, cai2022degradation}. However, these approaches are not directly compatible with DSSI due to its distinct optical priors. 

Existing DSSI studies \cite{xu2024video, lv2023aperture} typically benchmarked the networks originally developed to estimate HSIs from clear RGB inputs. The problem of these methods is the absence of explicit physical model guidance. Jeon's study \cite{jeon2019compact} and CSST \cite{lv2023aperture} have attempted to incorporate deep unfolding strategies. However, the former employs a gradient descent method (GDM) to solve the data fidelity subproblem, while the latter entirely replaces the fidelity solver with neural network layers. Without an analytical solution to the DSSI-specific data fidelity subproblem, these methods lack principled physical constraints, leading to suboptimal reconstruction performance.

As noted in \cite{chan2016plug}, the common challenge applying deep unfolding is whether one can obtain a fast solver for the inversion step of the data fidelity term. In the case of DSSI, the absence of such a solver poses a significant barrier to effective iterative reconstruction. Moreover, we observe that the iterative process is unstable when using non-learning-based methods or linear layers to generate initial input, leading to poor performance.

To address these issues, we derive an efficient analytical solution to the data fidelity subproblem of DSSI by leveraging transformations between the spatial and frequency domains, achieving an efficient analytical solver for the data fidelity term. Based on this, we propose a novel reconstruction framework for DSSI, termed the diffractive deep unfolding (DDU) framework. In our approach, a neural network is used to generate a suitable initialization for the iterative process. Both the initial estimate and the final output are supervised to ensure stable optimization. Our approach bridges traditional iterative model-based methods with modern learning-based approaches, enabling our framework to incorporate both the imaging model and the data-driven priors. Compared to previous DSSI deep unfolding methods, our approach significantly enhances the reconstruction performance. The main distinctions between our DDU with previous CASSI (DAUHST \cite{lv2023aperture}) and DSSI (Jeon's \cite{jeon2019compact}) unfolding methods are summarized in Table~\ref{tab: diff}.

\begin{table}[htpb]
\centering
\footnotesize
\caption{Difference between DDU, DAUHST, and Jeon's.}
\begin{tabular}{cccc}
\hline
  & DDU  & DAUHST & Jeon's \\ \hline
Fidelity Solver & \begin{tabular}[c]{@{}c@{}}Analytical\\ for DSSI\end{tabular}  & \begin{tabular}[c]{@{}c@{}}Analytical\\ for CASSI\end{tabular} & \begin{tabular}[c]{@{}c@{}}GDM\\ for DSSI\end{tabular}\\ 
Initialization & Network  & Conv & $\Phi^\mathsf{T}\mathrm{J}$ \\ 
Unfold Strategy  & ADMM & HQS & HQS \\
Params $\gamma,\widetilde\sigma,\zeta$ & Learned & Network & Learned \\
\hline
\end{tabular}
\label{tab: diff}
\end{table}

Extensive experiments demonstrate that our DDU is highly compatible with existing networks and enhances their performance. On the test dataset, DDU achieves superior reconstruction accuracy compared to the SOTA network with comparable parameter counts (Params) and floating point operations per second (FLOPS). In real-world experiments, it also achieved improved spectral accuracy and spatial resolution. Our contributions can be summarized as follows:
\begin{itemize}
\item We propose an efficient deep unfolding framework, DDU, for DSSI systems, which incorporates reliable physics-based priors to guide and enhance learning-based reconstruction networks.
\item We derive a fast and effective analytical solver for the data fidelity subproblem in DSSI. To improve stability, we use a neural network to initialize the iterative process and apply supervision to both the initial estimate and the final output to ensure robust optimization.
\item The proposed framework is highly compatible with existing network architectures. Both simulation and real-world experiments demonstrate that DDU consistently improves the performance of SOTA methods.

\end{itemize}
\section{Related Work}

\subsection{HSI Reconstruction}

Hyperspectral images can be reconstructed from compressed sensing. The NTIRE challenges have introduced many powerful networks \cite{li2020adaptive, zhao2020hierarchical, cai2022mst++} that estimate HSIs from clear RGB images. However, these methods suffer from the metamerism problem \cite{fu2024limitations}. Research on CASSI systems has also explored many end-to-end \cite{meng2020end, hu2022hdnet, cai2022mask} and deep unfolding \cite{wang2020dnu, cai2022degradation} methods. However, the complexity of CASSI systems limits their application outside the laboratory.
As a recently developed encoding approach, the DSSI system offers a more compact design using DOEs. Jeon et al. \cite{jeon2019compact} proposed a rotation diffracted system with a single DOE, reconstructed via GDM and UNet within a deep unfolding framework. Xu et al. \cite{xu2024video} extended this design by integrating a rear lens to balance PSF size, light throughput, and back focal length, and benchmarked networks originally developed for RGB-to-HSI estimation. Lv et al. \cite{lv2023aperture} designed a diffractive mask at the aperture and applied deep unfolding without an explicit data fidelity term. Several studies \cite{baek2021single, zhuge2025generalized} have focused on jointly optimizing DOE design and end-to-end reconstruction networks. More recently, quantization-aware methods \cite{li2022quantization, wang2024non} have been proposed to mitigate the impact of DOE quantization during processing. Although CASSI has inspired many physically guided reconstruction networks, these methods are typically incompatible with DSSI due to the fundamentally different optical priors. Reconstruction algorithms that explicitly incorporate the physical model of DSSI remain underexplored. 
This work aims to fill this gap.

\subsection{Deep Unfolding Network}

Conventional model-based methods, e.g. alternating direction method of multipliers (ADMM), typically separate data fidelity and prior terms, leading to an iterative process alternating between solving a data fidelity subproblem and a prior subproblem. Deep unfolding networks reformulate the prior subproblem as a denoising task and address it using neural networks. This approach has been applied to image super-resolution \cite{zhang2020deep}, magnetic resonance compressive imaging \cite{sun2016deep}, video snapshot compressive imaging \cite{ma2019deep} and CASSI \cite{wang2020dnu}. However, these unfolding methods are not directly compatible with DSSI systems due to differences in their sensing matrices, which significantly affect the solver of the data fidelity subproblem. Image super-resolution involves a one-channel to one-channel deconvolution problem, for which Zhao et al.\cite{zhao2016fast} proposed a closed-form solver. CASSI reconstruction involves a one-channel to multi-channel mapping, with a corresponding solver introduced by Yuan\cite{yuan2016generalized}. In contrast, DSSI reconstruction requires solving a three-channel to multi-channel deconvolution problem, for which no analytical solution has previously been established. Although Jeon et al.\cite{jeon2019compact} and Lv et al.\cite{lv2023aperture} have attempted to adapt deep unfolding to DSSI, both approaches bypassed the derivation of a closed-form solution for the data fidelity term, resulting in suboptimal reconstruction performance.
In this work, we derive an analytical solution to the data fidelity subproblem in DSSI and leverage the structure of diagonal matrices for efficient implementation, significantly enhancing the effectiveness of the unfolding process. Furthermore, we carefully design the initialization strategy and loss supervision to enable the unfolding framework to improve the performance of existing state-of-the-art networks. We also conduct comprehensive ablation studies to determine the optimal configuration of our framework structure.

\section{Method}
\subsection{Optical Encoding and Image Degradation}

\begin{figure}[htpb]
  \centering
  \includegraphics[width=\linewidth]{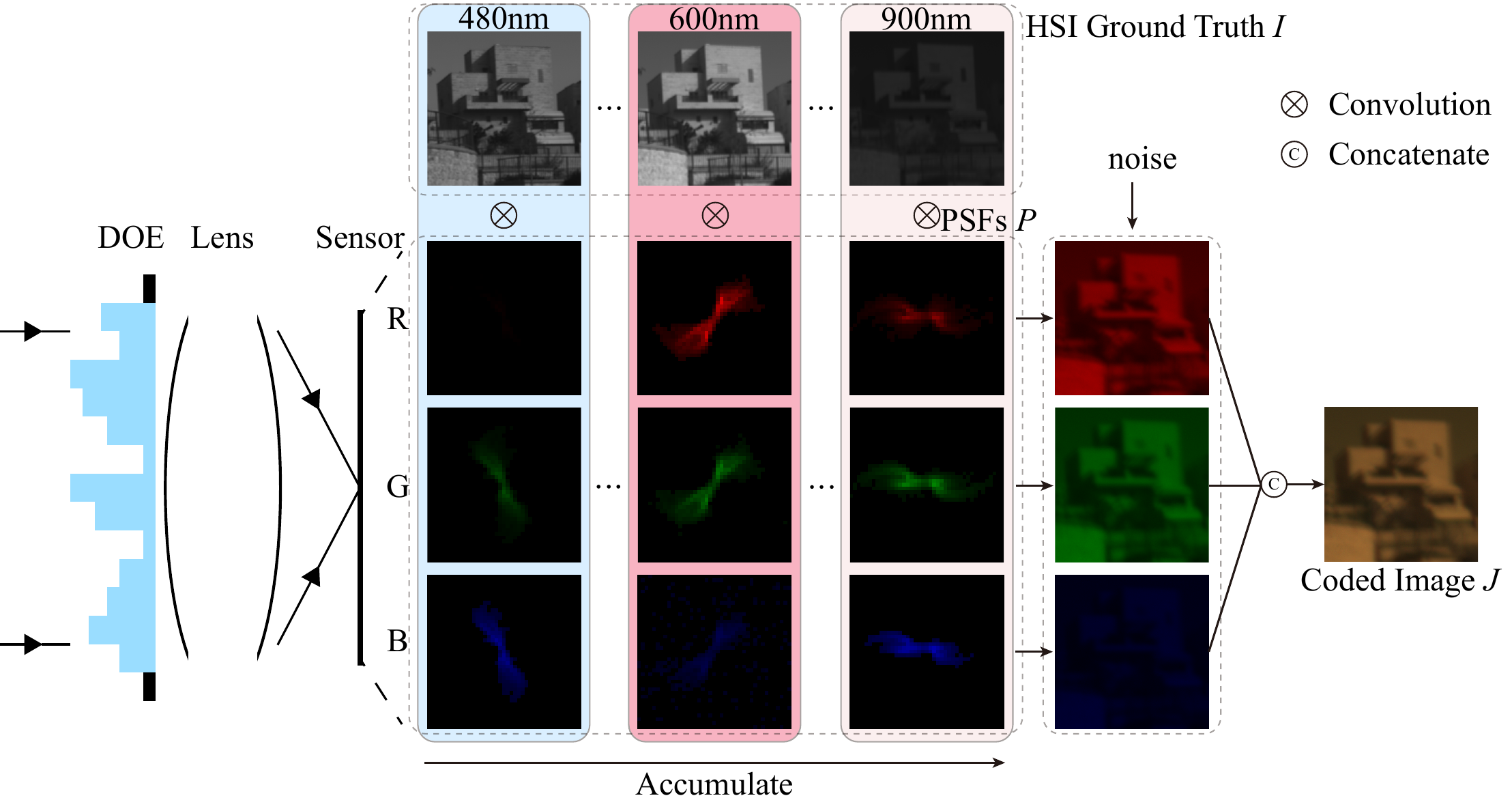}
  \caption{The encoding mechanism of DSSI system.}
  \label{fig: Encoding}
\end{figure}

The DOE design of the DSSI system has been extensively studied in prior work, primarily to ensure that incident waves of different wavelengths produce distinct PSFs. The mechanism of spectral encoding through PSFs is illustrated in Fig.~\ref{fig: Encoding}.

For a spectral image $I[x,y,i]$, the modulation of a DSSI system in discrete form with normalized PSFs $p[x,y,i]$ and spectral response function $\Omega[c,i]$ at color channel $c\in \{r,g,b\}$ can be expressed as:
\begin{equation}
    \begin{aligned}
        J[x,y,c] &= \sum_{i=1}^{N_\lambda} I[x,y,i] \otimes p[x,y,i] \cdot \Omega[c,i] + \eta,\\
                &=\sum_{i=1}^{N_\lambda} I[x,y,i] \otimes P[x,y,c,i] + \eta,
    \end{aligned}
    \label{eq: degrade}
\end{equation}
where $i$ denotes the $i^{th}$ spectral channel, $N_\lambda$ is the total number of spectral channels, $\eta$ is imaging noise, $J[x,y,c]$ is the coded image at color channel $c$, $\otimes$ is defined as the 2D convolution operator and $P[x,y,c,i] = p[x,y,i] \cdot \Omega[c,i]$ are the unified PSFs. 
The simulated image degradation in this paper is performed in the spatial domain, following Eq.~\ref{eq: degrade}.

Let $\mathbf{I}\in\mathbb{R}^{nN_\lambda}$ be the original HSI vector and $\mathbf{J}\in \mathbb{R}^{3n}$ be the coded RGB image vector, where $n$ is the product of the image width $W$ and height $H$. Eq.~\ref{eq: degrade} can be further written into vectorized form:
\begin{equation}
    \mathbf{J} =\boldsymbol{\Phi} \mathbf{I} + \mathbf{n},
    \label{eq: vectorized_degrade}
\end{equation}
where $\boldsymbol{\Phi} \in \mathbb{R}^{3n\times nN_\lambda}$ is the convolution matrix and $\mathbf{n}$ is the vectorized imaging noise.

Eq.~\ref{eq: vectorized_degrade} can be transferred to the frequency domain with the circular boundary assumption. Given the vectorized spectral image spectrum $\mathbf{U}\in\mathbb{C}^{nN_\lambda}$ and the measure matrix $\mathbf{H}\in\mathbb{C}^{3n \times nN_\lambda}$, the coded image spectrum $\mathbf{V}\in\mathbb{C}^{3n}$ can be represented by the degradation model as
\begin{equation}
    \mathbf{V} = \mathbf{H}\mathbf{U}.
    \label{eq: vectorized_degrade_frequency}
\end{equation}

For clarity, $\mathbf{V}$, $\mathbf{U}$, and $\mathbf{H}$ can be written as block diagonal matrices:
\begin{equation}
\begin{aligned}
\mathbf{V}&=\begin{bmatrix}\mathbf{V}_{r}&\mathbf{V}_{g}&\mathbf{V}_{b}\end{bmatrix}^\mathsf{T},\\
\mathbf{U}&=\begin{bmatrix}\mathbf{U}_{1}&\mathbf{U}_{2}&\cdot\cdot\cdot&\mathbf{U}_{N_\lambda}\end{bmatrix}^\mathsf{T},\\
\mathbf{H}&=\begin{bmatrix}
\mathbf{H}_{r1}&\mathbf{H}_{r2}&\cdot\cdot\cdot&\mathbf{H}_{rN_\lambda}\\
\mathbf{H}_{g1}&\mathbf{H}_{g2}&\cdot\cdot\cdot&\mathbf{H}_{gN_\lambda}\\
\mathbf{H}_{b1}&\mathbf{H}_{b2}&\cdot\cdot\cdot&\mathbf{H}_{bN_\lambda}\\
\end{bmatrix},
\end{aligned}
\end{equation}
where the diagonal elements of $\mathbf{V}_c \in \mathbb{C}^{n\times n}$, $\mathbf{U}_i \in \mathbb{C}^{n\times n}$ and $\mathbf{H}_{ci}\in \mathbb{C}^{n\times n}$ correspond to the 1-D representation of $\mathcal{F}(J[x,y,c])$, $\mathcal{F}(I[x,y,i])$ and $\mathcal{F}(P[x,y,c,i])$. The function $\mathcal{F}(\cdot)$ denotes the Fourier transform.

\subsection{Deep Unfolding Framework}

\begin{figure}[htpb]
  \includegraphics[width=1.0\linewidth]{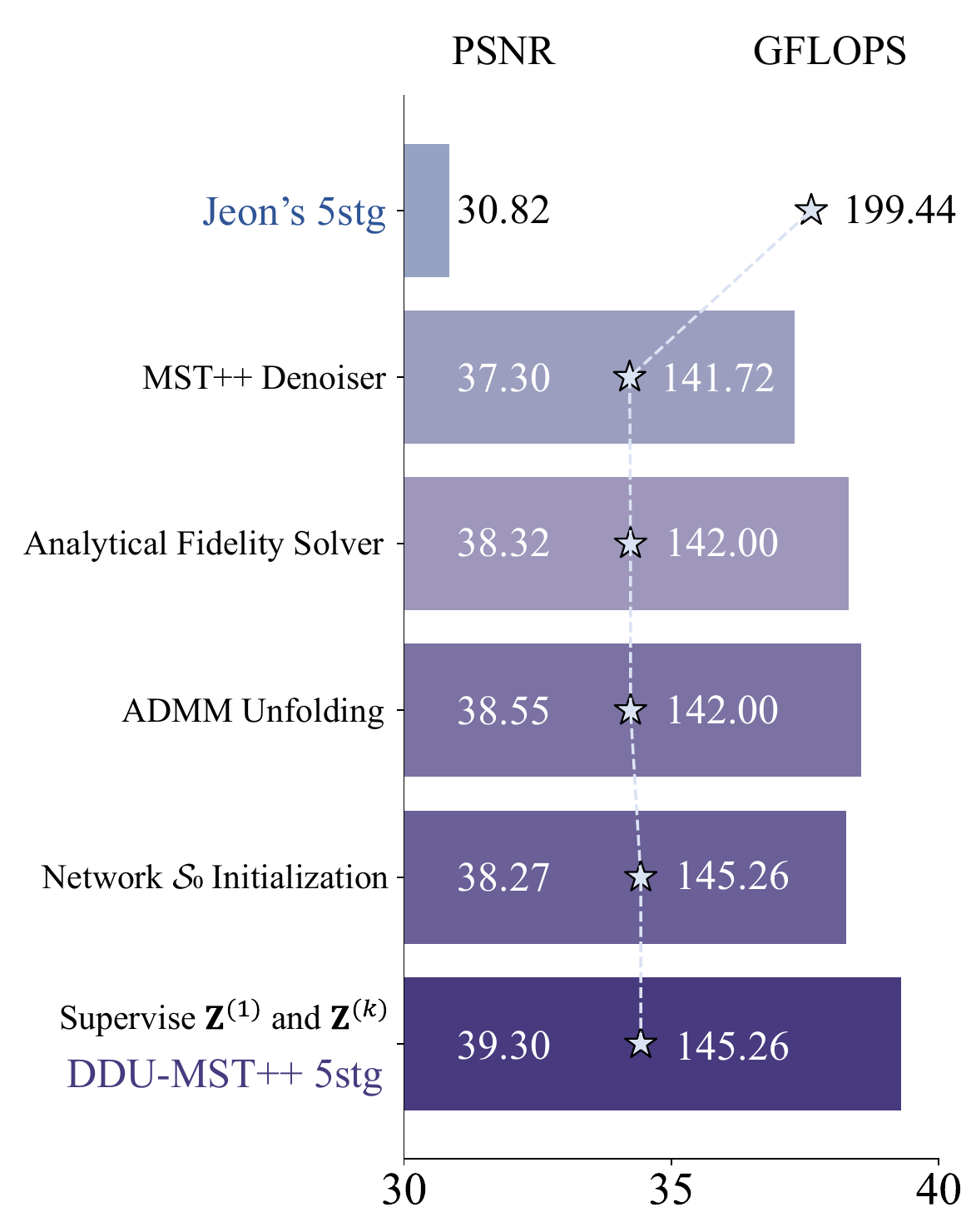}
  \caption{Evolution from Jeon's 5stg method to DDU-MST++ 5stg.}
  \label{fig: modern}
\end{figure}

\begin{figure*}
  \centering
  \includegraphics[width=\linewidth]{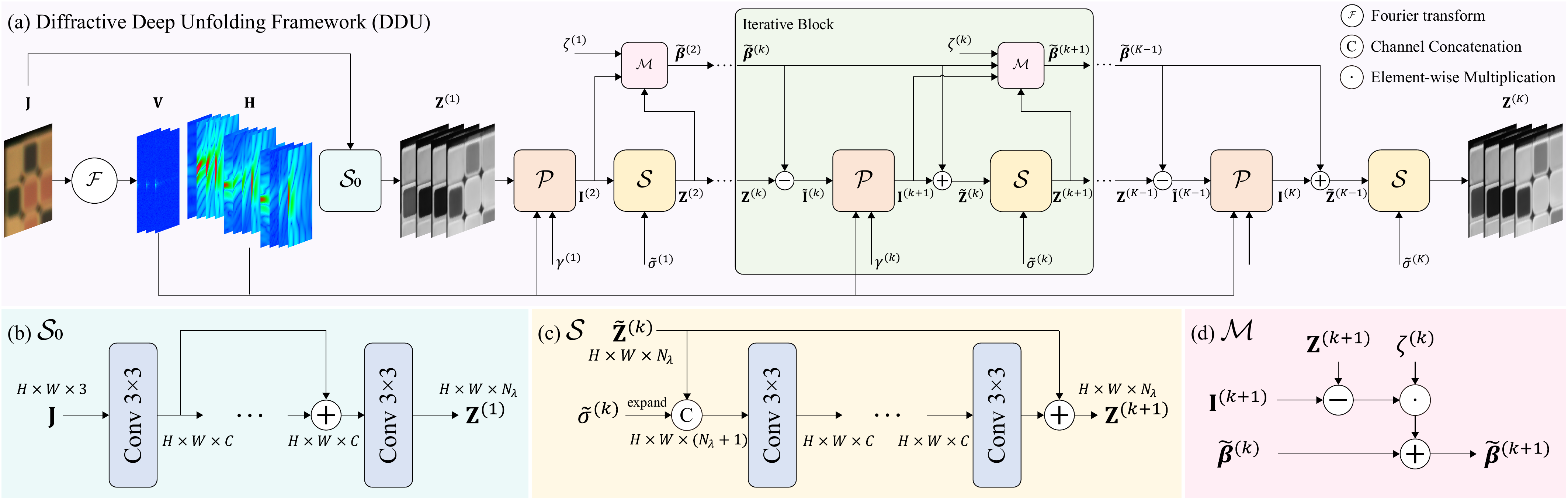}
  \caption{(a) The overview of our DDU. $\mathcal{S}_0$ is the initialization network. $\mathcal{P}$, $\mathcal{S}$ and $\mathcal{M}$ denote the linear projection operator, nonlinear denoising network and Lagrangian multipliers updater. (b) Abstracted structure of $\mathcal{S}_0$. (c) Abstracted structure of $\mathcal{S}$. (d) Structure of $\mathcal{M}$.}
  \label{fig: Chain}
\end{figure*}

The pioneering work of diffracted rotation \cite{jeon2019compact} attempted to use deep HQS to extract HSI from Eq.~\eqref{eq: vectorized_degrade}. However, the matrix involved is highly complex, making it infeasible to compute its inverse directly. They used GDM to bypass matrix inversion, but this approach significantly slowed the convergence process.

In Fig.~\ref{fig: modern}, we illustrate the step-by-step evolution from Jeon's original 5-stage method to our proposed 5-stage DDU-MST++. This modernization process involves the following key modifications:
(1) replacing Jeon's original denoiser with MST++;
(2) substituting GDM with our analytical data fidelity solver and switching from HQS-based unfolding to ADMM-based unfolding;
(3) introducing a learnable initialization network $\mathcal{S}_0$ to generate the initial input for the iterative process;
(4) applying supervision to both the initialization $\mathbf{Z}^{(1)}$ and the final reconstruction result $\mathbf{Z}^{(K)}$.

We first tackle the core issue by deriving an efficient analytical solution to the data fidelity subproblem in the frequency domain. The original spectral image spectrum can be estimated by solving the following minimization problem:

\begin{equation}
    \widehat{\mathbf{I}}=\mathop {\arg \min }\limits _{\mathbf{I}} {\frac{1}{2}} \Vert\boldsymbol{\Phi} \mathbf{I}-\mathbf{J}\Vert ^{2}_{2}+\sigma R(\mathbf{I}),
\end{equation}
where $R(\cdot)$ is the prior term with some priors and $\sigma$ is a parameter that tweaks the weights of the data fidelity term and the prior term. We use ADMM to solve this problem. By introducing an auxiliary variable $\mathbf{Z}$, it can be formulated as an unconstrained optimization problem:
\begin{equation}
    \widehat{\mathbf{I}}=\mathop {\arg \min }\limits _{\mathbf{I}}  {\frac{1}{2}} \Vert\boldsymbol{\Phi} \mathbf{I}-\mathbf{J}\Vert ^{2}_{2}+\sigma R(\mathbf{Z}), \ s.t.\ \mathbf{Z}=\mathbf{I},
\end{equation}
and consider its augmented Lagrangian function
\begin{equation}
    \mathcal{L}(\mathbf{I}, \mathbf{Z}, \boldsymbol{\beta})=\frac{1}{2}\Vert \boldsymbol{\Phi} \mathbf{I}-\mathbf{J}\Vert ^{2}_{2}+ \frac{\gamma}{2}\Vert\mathbf{Z}-\mathbf{I}\Vert_2^2 + \boldsymbol{\beta}^\mathsf{T}(\mathbf{Z}-\mathbf{I}) + \sigma R(\mathbf{Z}),
    \label{eq: Lagrangian_deblur}
\end{equation}
where $\gamma$ is a penalty parameter and $\boldsymbol{\beta}\in \mathbb{R}^{n}$ are Lagrangian multipliers.
Eq.~\ref{eq: Lagrangian_deblur} can be splitted into three iterative subproblems:
\begin{equation}
    \mathbf{I}^{(k+1)}=\mathop {\arg \min }\limits _{\mathbf{I}} \frac{1}{2}\Vert\boldsymbol{\Phi}\mathbf{I}-\mathbf{J}\Vert_2^2+\frac{\gamma^{(k)}}{2}\Vert\mathbf{I}-\widetilde{\mathbf{I}}^{(k)}\Vert_2^2,
    \label{eq: iter_I}
\end{equation}
\begin{equation}
    \mathbf{Z}^{(k+1)}=\mathop {\arg \min }\limits _{\mathbf{Z}}\frac{\gamma^{(k)}}{2} \Vert\mathbf{Z}-\widetilde{\mathbf{Z}}^{(k)}\Vert_2^2+\sigma^{(k)} R(\mathbf{Z}),
    \label{eq: iter_Z}
\end{equation}
\begin{equation}
    \widetilde{\boldsymbol{\beta}}^{(k+1)}=\widetilde{\boldsymbol{\beta}}^{(k+1)}+\zeta^{(k)}(\mathbf{I}^{(k+1)}-\mathbf{Z}^{(k+1)}),
    \label{eq: iter_beta}
\end{equation}
where $\zeta^{(k)}$ is the update rate, $\widetilde{\boldsymbol{\beta}}^{(k)}\stackrel{\text{def}}{=}\boldsymbol{\beta}^{(k)}/\gamma^{(k)}$, $\widetilde{\mathbf{I}}^{(k)}\stackrel{\text{def}}{=}\mathbf{Z}^{(k)}-\widetilde{\boldsymbol{\beta}}^{(k)}$ and $\widetilde{\mathbf{Z}}^{(k)}\stackrel{\text{def}}{=}\mathbf{I}^{(k+1)}+\widetilde{\boldsymbol{\beta}}^{(k)}$. 
It can be observed that the data fidelity term and the prior term are decoupled into the data fidelity subproblem Eq.~\ref{eq: iter_I} and the prior subproblem Eq.~\ref{eq: iter_Z}, respectively. Eq.~\ref{eq: iter_I} represents a three-channel to multi-channel deconvolution problem, which is highly ill-posed. To address this, we transform the problem into the frequency domain:
\begin{equation}
    \mathbf{U}^{(k+1)}=\mathop {\arg \min }\limits _{\mathbf{U}} \frac{1}{2}\Vert\mathbf{H}\mathbf{U}-\mathbf{V}\Vert_2^2+\frac{\gamma^{(k)}}{2}\Vert\mathbf{U}-\widetilde{\mathbf{U}}^{(k)}\Vert_2^2,
    \label{eq: iter_U}
\end{equation}
where $\widetilde{\mathbf{U}}^{(k)}$ is the frequency representation of $\widetilde{\mathbf{I}}^{(k)}$.

Eq.~\ref{eq: iter_U} has a closed-form solution as 
\begin{equation}
    \begin{aligned}
        \mathbf{U}^{(k+1)}&=(\mathbf{H}^*\mathbf{H} +\gamma^{(k)} \mathbf{E})^{-1}(\mathbf{H}^*\mathbf{V}+\gamma^{(k)} \widetilde{\mathbf{U}}^{(k)})\\
        &=\widetilde{\mathbf{U}}^{(k)}+ 
        \frac{1}{\gamma^{(k)}}\mathbf{H}^*[\mathbf{V}-\\
        &\mathbf{A}^{-1}(\frac{1}{\gamma^{(k)}}\mathbf{H}\mathbf{H}^*\mathbf{V}+\mathbf{H}\widetilde{\mathbf{U}}^{(k)})],
    \end{aligned}
    \label{eq: solve_U}
\end{equation}
where $\mathbf{A}=\mathbf{E}+\frac{1}{\gamma^{(k)}}\mathbf{H}\mathbf{H}^*$. As $\mathbf{H}$ consists of $3 \times N_\lambda$ diagonal blocks, $\mathbf{H}\mathbf{H}^*\in\mathbb{C}^{n\times n}$ and $\mathbf{A}$ are block diagonal matrices as well.
The inversion of the third-order block matrix $\mathbf{A}$ can be calculated recursively. Let $\mathbf{G}=\frac{1}{\gamma^{(k)}}\mathbf{H}\mathbf{H}^*$, which is a block diagonal matrix. The inversion of $\mathbf{A}$ can be written as

\begin{equation}
    \begin{aligned}
    \mathbf{A}^{-1}
    &=\begin{bmatrix}
    \mathbf{G}_{11}+\mathbf{E}&\mathbf{G}_{12}&\mathbf{G}_{13}\\
    \mathbf{G}_{21}&\mathbf{G}_{22}+\mathbf{E}&\mathbf{G}_{23}\\
    \mathbf{G}_{31}&\mathbf{G}_{32}&\mathbf{G}_{33}+\mathbf{E}\\
    \end{bmatrix}^{-1}\\
    &=\begin{bmatrix}
    \mathbf{A}_{11}&\mathbf{A}_{12}&\mathbf{A}_{13}\\
    \mathbf{A}_{21}&\mathbf{A}_{22}&\mathbf{A}_{23}\\
    \mathbf{A}_{31}&\mathbf{A}_{32}&\mathbf{A}_{33}\\
    \end{bmatrix}^{-1},
    \end{aligned}
\end{equation}
where each block matrix is a diagonal matrix. Define
\begin{equation}
    \begin{aligned}
        \mathbf{B}_{11}=\begin{bmatrix}
        \mathbf{A}_{11}&\mathbf{A}_{12}\\
        \mathbf{A}_{21}&\mathbf{A}_{22}
        \end{bmatrix}&,
        \mathbf{B}_{12}=\begin{bmatrix}
        \mathbf{A}_{13}\\
        \mathbf{A}_{23}
        \end{bmatrix},\\
        \mathbf{B}_{21}=\begin{bmatrix}
        \mathbf{A}_{31}&\mathbf{A}_{32}
        \end{bmatrix}&,
        \mathbf{B}_{22}=
        \mathbf{A}_{33}.
    \end{aligned}
\end{equation}
Then $\mathbf{A}$ can be seen as a second-order block matrix and its inversion is
\begin{equation}
    \begin{aligned}
        \mathbf{A}^{-1}=&\begin{bmatrix}
        \mathbf{B}_{11}&\mathbf{B}_{12}\\
        \mathbf{B}_{21}&\mathbf{B}_{22}
        \end{bmatrix}^{-1}\\
        =&\begin{bmatrix}
        \mathbf{B}_{11}^{-1}+\mathbf{B}_{11}^{-1}\mathbf{B}_{12}\mathbf{D}\mathbf{B}_{21}\mathbf{B}_{11}^{-1}&
        -\mathbf{B}_{11}^{-1}\mathbf{B}_{12}\mathbf{D}\\
        -\mathbf{D}\mathbf{B}_{21}\mathbf{B}_{11}^{-1}&
        \mathbf{D}
        \end{bmatrix},
    \end{aligned}
\end{equation}
where $\mathbf{D}=(\mathbf{B}_{22}-\mathbf{B}_{21}\mathbf{B}_{11}^{-1}\mathbf{B}_{12})^{-1}$ is a diagonal matrix. The inversion of the second-order block matrix $\mathbf{B}_{11}$ can be written as 
\begin{equation}
    \mathbf{B}_{11}^{-1}=
    \begin{bmatrix}
    \mathbf{A}_{11}^{-1}+\mathbf{A}_{11}^{-1}\mathbf{A}_{12}\mathbf{C}\mathbf{A}_{21}\mathbf{A}_{11}^{-1}&
    -\mathbf{A}_{11}^{-1}\mathbf{A}_{12}\mathbf{C}\\
    -\mathbf{C}\mathbf{A}_{21}\mathbf{A}_{11}^{-1}&
    \mathbf{C}
    \end{bmatrix},
\end{equation}
where $\mathbf{C}=(\mathbf{A}_{22}-\mathbf{A}_{21}\mathbf{A}_{11}^{-1}\mathbf{A}_{12})^{-1}$ is a diagonal matrix. 

Due to the block diagonal nature of these matrices, $\mathbf{A}^{-1}$ and Eq.~\ref{eq: solve_U} can be further simplified to calculate diagonal elements, significantly reducing computational complexity. This simplification makes it possible to calculate $\mathbf{A}^{-1}$ at each iteration and optimize $\gamma$ during training efficiently.

In particular, Eq.~\ref{eq: iter_I} can be regarded as an inversion step, since it involves the degradation model. The parameter $\gamma^{(k)}$ controls the balance between the data fidelity and the prior. When $\gamma^{(k)}$ is small enough, the output $\mathbf{I}^{(k+1)}$ is primarily constrained by the degradation model and can be reverted to the noisy coded image $\mathbf{J}$ via $\mathbf{\Phi}$, although the output remains noisy. In contrast, when $\gamma^{(k)}$ is large enough, the output $\mathbf{I}^{(k+1)}$ closely approximates $\widetilde{\mathbf{I}}^{(k)}$, which incorporates prior information. Therefore, $\gamma^{(k)}$ should be small during the early stages and large in the final stage to fully exploit the data fidelity term and ensure the convergence to a fixed point. 

For Eq.~\ref{eq: iter_Z}, it can be seen as a nonlinear denoising step:
\begin{equation}
        \mathbf{Z}^{(k+1)}=\mathop {\arg \min }\limits _{\mathbf{Z}}\frac{1}{2(\sqrt{\sigma^{(k)}/\gamma^{(k)}})^2} \Vert\mathbf{Z}-\widetilde{\mathbf{Z}}^{(k)}\Vert_2^2+ R(\mathbf{Z}).
        \label{eq: iter_Z_denoise}
\end{equation}
It is equivalent to denoise $\widetilde{\mathbf{Z}}^{(k)}$ with noise level at $\widetilde{\sigma}^{(k)}=\sqrt{\sigma^{(k)}/\gamma^{(k)}}$ using prior $R(\mathbf{Z})$\cite{chan2016plug}.

As shown in Fig.~\ref{fig: Chain}, we formulate the decoding process as an iterative scheme:
\begin{equation}
    \begin{aligned}
        \mathbf{I}^{(k+1)}&=\mathcal{P}(\mathbf{J},\widetilde{\mathbf{I}}^{(k)},\gamma^{(k)},\mathbf{H}),\\
        \mathbf{Z}^{(k+1)}&=\mathcal{S}(\widetilde{\mathbf{Z}}^{(k)},\widetilde{\sigma}^{(k)}),\\ 
        \widetilde{\boldsymbol{\beta}}^{(k+1)}&=\mathcal{M}(\widetilde{\boldsymbol{\beta}}^{(k)},\mathbf{I}^{(k+1)},\mathbf{Z}^{(k+1)},\zeta^{(k)}),
    \end{aligned}
\end{equation}
where $\mathcal{P}$ is the linear projection operator defined in Eq.~\ref{eq: solve_U}, $\mathcal{S}$ is the nonlinear projection that solves Eq.~\ref{eq: iter_Z_denoise}, and $\mathcal{M}$ updates the Lagrangian multipliers as in Eq.~\ref{eq: iter_beta}.
The $\widetilde{\sigma}^{(k)}$, $\gamma^{(k)}$, and $\zeta^{(k)}$ are parameters optimized during training. 

Since the problem is highly ill-posed, the stability of the iterative process is strongly affected by the quality of the initial input. Conventional strategies, such as $\boldsymbol{\Phi}^\mathsf{T}\mathbf{J}$ or a learnable linear layer, often lead to oscillatory behavior during iteration. To address this, our framework initializes $\mathbf{Z}^{(1)}$ from $\mathbf{J}$ using a dedicated neural network, denoted $\mathcal{S}_0$.


We do not specifically design $\mathcal{S}_0$ and $\mathcal{S}$. Instead, we modify existing cascade networks (MST++ and MIRNetV2) for ease of comparison, whose main structure is based on stacking the same block. The abstracted structures of $\mathcal{S}_0$ and $\mathcal{S}$ are illustrated in Fig.~\ref{fig: Chain}b and Fig.~\ref{fig: Chain}c. Our goal is to integrate these networks into our framework with minimal modifications. The embedding dimensions $C$ of $\mathcal{S}_0$ and $\mathcal{S}$ remain the same, set to 64 and 80 when we adopt MST++ and MIRNetV2, respectively. 
The weights of $\mathcal{S}$ are not shared across different stages, and $\widetilde{\sigma}$ is disabled.
\begin{figure*}[ht]
  \centering
  \includegraphics[width=1.0\linewidth]{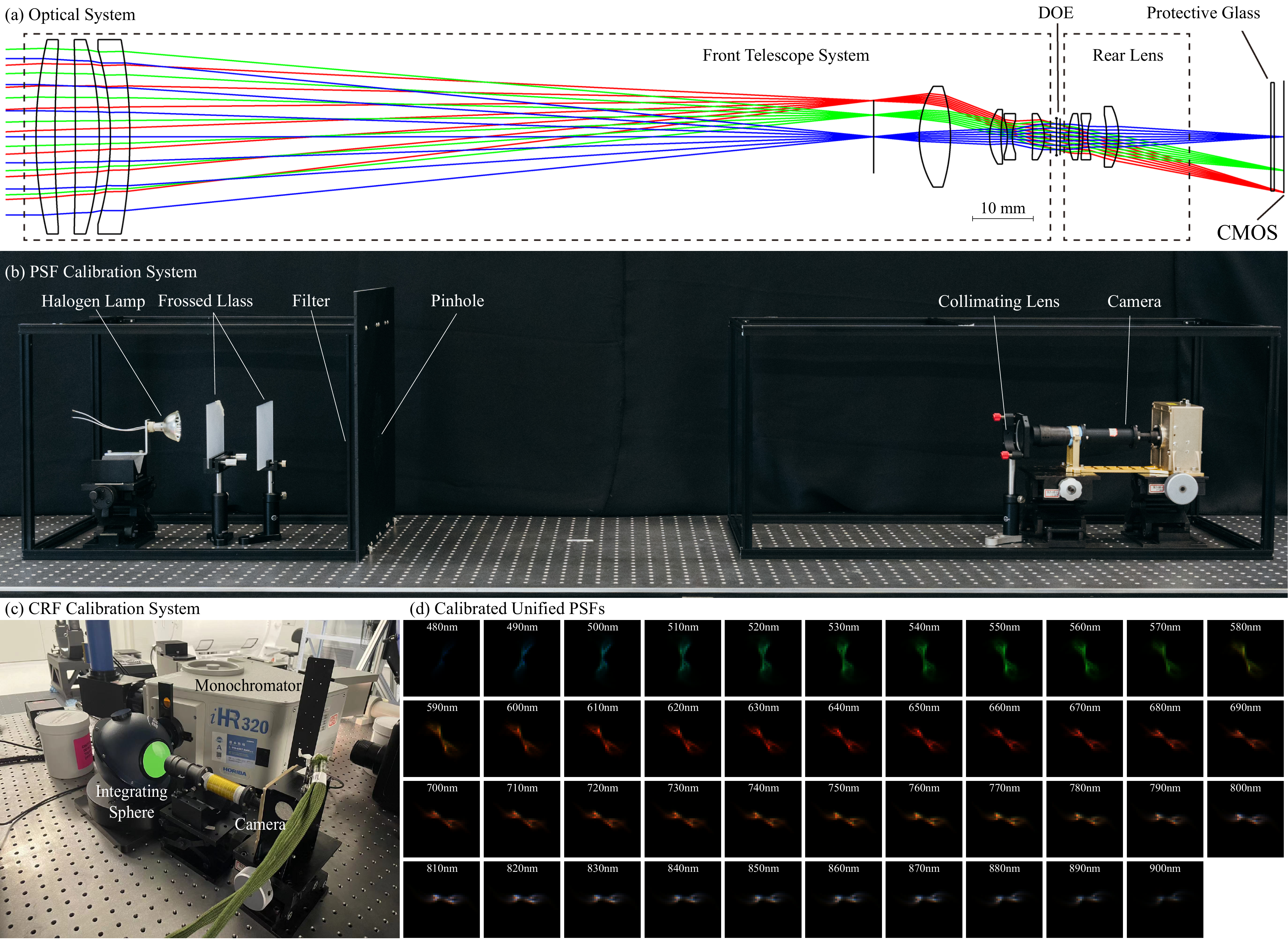}
  \caption{(a) The optical structure of the DSSI system. (b) Point spread function calibration system. (c) Camera response function calibration system. (d) The calibrated PSFs $P$ at different wavelengths. The size of each PSF is $41\times41\times3$.}
  \label{fig: optics}
\end{figure*}

We train networks with $K$ stages using mean absolute error (MAE) and structural similarity (SSIM) \cite{wang2004image} loss:
\begin{equation}
    \mathcal L = \sum_{k\in \{1,K\}}\beta(1 - SSIM(\mathbf{Z}^{(k)}, \mathbf{I})) + (1-\beta)||\mathbf{Z}^{(k)}-\mathbf{I}||,
\end{equation}
where $\beta$ is set to 0.85 empirically \cite{zhao2016loss}. Notably, we supervise both the initial estimate $\mathbf{Z}^{(1)}$ and the final output $\mathbf{Z}^{(K)}$. Supervising only the final output $\mathbf{Z}^{(K)}$ leads to suboptimal performance, as illustrated in Fig.~\ref{fig: modern} and further validated through our ablation study.

\section{Experiments}

\subsection{Hardware and PSF Calibration}

In this paper, we adopt a diffracted rotation design with two wings covering the spectral range from 480 nm to 900 nm. Our optical configuration follows that of Xu et al. \cite{xu2024video}, as shown in Fig.~\ref{fig: optics}a. The diffractive optical element (DOE) has a focal length of 150 mm and a diameter of 5 mm, while the rear lens group has a focal length of 42.3 mm. The telescope extends the field of view, and the rear lens reduces the PSF size and shortens the back focal distance. These two components are not essential for a DSSI system. Without the telescope, the system length is less than 5 cm, making it highly compact. We calibrate the normalized PSFs $p$ and measure the spectral response function $\Omega$ using calibration systems shown in Fig.~\ref{fig: optics}b and Fig.~\ref{fig: optics}c. Then we get the unified PSFs shown in Fig.~\ref{fig: optics}d.

The sensor originally outputs 12-bit depth raw images. To reduce transmission bandwidth, the raw images are pixel-binned to three-channel RGB images. Within each $2\times2$ pixel region of the Bayer pattern, the values of the R and B pixels are retained as the R and B color channels, while the two G pixels are averaged to form the G channel.
To mitigate noise, 15 dark images are captured to remove stripe noise and reduce dark noise. For scene capture, 5 consecutive frames are averaged as the input for reconstruction. According to our calibration results, the imaging noise $\eta$ with 1.5 ms exposure can be modeled as Gaussian noise with a standard deviation of 7e-5 and 14-bit Poisson noise.

\begin{table*}[htpb]
\centering
\caption{Quantitative results on ICVL \cite{arad2016sparse} and VNIRSR \cite{xu2025two} datasets.}
\begin{tabular}{ccccccccc}
\hline
Method         & Params & GFLOPS & \multicolumn{2}{c}{PSNR $\uparrow$} & \multicolumn{2}{c}{SAM $\downarrow$} & \multicolumn{2}{c}{SSIM $\uparrow$} \\ \cline{4-9} 
                  &        &        & ICVL  &  VNIRSR & ICVL   & VNIRSR  & ICVL   & VNIRSR  \\ \hline
EDSR \cite{lim2017enhanced} & 2.43M & 158.77 & 35.98 & 29.64 & 0.0832 & 0.0950 & 0.9613 & 0.9288 \\
HDNet \cite{hu2022hdnet}    & 2.36M & 154.37 & 39.95 & 34.60 & 0.0531 & 0.0569 & 0.9699 & 0.9478 \\
MPRNet \cite{zamir2021multi} & 12.04M & 421.26 & 42.30 & 37.39 & 0.0334 & 0.0409 & 0.9765 & 0.9592 \\
MambaIR \cite{guo2024mambair} & 1.55M & 156.10 & 41.64 & 35.85 & 0.0375 & 0.0482 & 0.9747 & 0.9543 \\ 
MIRNetV2 1stg \cite{zamir2022learning} & 2.45M  & 65.52 & 41.35 & 35.59 & 0.0388 & 0.0497 & 0.9737 & 0.9509 \\
MIRNetV2 3stg \cite{zamir2022learning} & 7.30M  & 192.23 & 42.09 & 36.16 & 0.0448 & 0.0453 & 0.9750 & 0.9547 \\
MIRNetV2 5stg \cite{zamir2022learning} & 12.14M  & 318.93 & 42.62 & 36.40 & 0.0384 & 0.0439 & 0.9761 & 0.9561 \\
MIRNetV2 7stg \cite{zamir2022learning} & 16.98M & 445.64 & 42.32 & 37.12 & 0.0340 & 0.0431 & 0.9763 & 0.9575 \\
MST++ 1stg \cite{cai2022mst++} & 2.28M  & 31.67  & 41.67 & 36.08 & 0.0372 & 0.0471 & 0.9748 & 0.9539 \\
MST++ 3stg \cite{cai2022mst++} & 6.77M  & 91.52  & 42.44 & 37.13 & 0.0326 & 0.0430 & 0.9766 & 0.9578 \\
MST++ 5stg \cite{cai2022mst++} & 11.27M & 151.38 & 42.50 & 37.32 & 0.0326 & 0.0419 & 0.9768 & 0.9597 \\
MST++ 7stg \cite{cai2022mst++} & 15.77M & 211.24 & 42.43 & 37.13 & 0.0329 & 0.0430 & 0.9771 & 0.9586 \\ \hline
ISTANet 9stg \cite{zhang2018ista} & 1.77M  & 159.68 & 31.85 & 25.52 & 0.1470 & 0.1585 & 0.9434 & 0.9064 \\
Jeon's 9stg \cite{jeon2019compact} & 69.71M & 358.99 & 35.54 & 28.89 & 0.0869 & 0.1012 & 0.9599 & 0.9223 \\
DPCDUN 5stg \cite{song2023dynamic} & 2.05M & 133.09 & 35.39 & 27.53 & 0.1031 & 0.1158 & 0.9560 & 0.9137 \\
CSST 5stg \cite{lv2023aperture} & 11.88M & 163.03 & 41.89 & 36.21 & 0.0348 & 0.0470 & 0.9750 & 0.9547 \\
DDU-MIRNetV2 3stg & 7.31M  & 192.91 & 42.34 & 36.54 & 0.0425 & 0.0438 & 0.9756 & 0.9561 \\
DDU-MIRNetV2 5stg & 12.16M  & 320.29 & 42.82 & 36.92 & 0.0372 & 0.0417 & 0.9770 & 0.9571 \\
DDU-MIRNetV2 7stg & 17.01M & 447.67 & 42.49 & 37.48 & 0.0326 & 0.0412 & 0.9765 & 0.9595 \\
DDU-MST++ 3stg    & 6.72M  & 88.47  & 42.59 & 37.46 & 0.0320 & 0.0411 & 0.9769 & 0.9587 \\
DDU-MST++ 5stg & 11.17M & 145.26 & 42.60            & 37.66            & 0.0317            & \textbf{0.0400}  & 0.9771           & 0.9602           \\
DDU-MST++ 7stg & 15.62M & 202.06 & \textbf{42.79}   & \textbf{37.70}   & \textbf{0.0311}   & 0.0401           & \textbf{0.9775}  & \textbf{0.9608}  \\ \hline
\end{tabular}
\label{tab: quantitative}
\end{table*}

\subsection{Experimental Setup}

In this paper, the ICVL \cite{arad2016sparse} and VNIRSR \cite{xu2025two} datasets are used to train the optimizable parameters. The ICVL dataset, captured using a commercial hyperspectral imager, contains 201 scenes with a spatial resolution of approximately $1300\times1392$ and spectral coverage from 400 to 1000 nm at intervals of approximately 1.25 nm. We resample it to the range of 480 to 900 nm with 10 nm intervals. The VNIRSR dataset, captured using the GaiaField-V10 HR hyperspectral imager, consists of 330 scenes with a spatial resolution of $936\times960$ and spectral coverage from 400 to 900 nm with 10 nm intervals. We select 125 and 300 scenes from ICVL and VNIRSR as the training set, and 15 and 30 scenes as the test set. We also define a mixed test set comprising samples from the ICVL and VNIRSR test sets, where the evaluation metric is calculated as a weighted sum: $\frac{1}{3} \text{ICVL}+ \frac{2}{3} \text{VNIRSR}$.

The training HSI is randomly $2\times$ downsampled and then randomly cropped into $336\times336$ patches. The coded images are then generated following Eq.~\ref{eq: degrade} and are centrally cropped with size of $296\times296$. Data augmentation techniques are applied, including random flipping, random contrast adjustments, and random metamers ($\alpha \in [-1,2]$) \cite{fu2024limitations}. To simulate real imaging situations, we add 14-bit Poisson noise. Additionally, Gaussian noise with a standard deviation of 5e-3 is injected to improve generalization. 
For testing, we use 2 fixed metamers ($\alpha \in \{0,1\}$) to augment the test set, where $\alpha = 1$ denotes the original data. We crop the central $936\times 952$ patches and degrade them as training.
The outputs are cropped 20 pixels at the edges before calculating the loss and evaluation metrics.


Our DDU framework is implemented using PyTorch. The parameters are optimized using AdamW \cite{loshchilov2017decoupled} optimizer with a cosine annealing scheduler for 30 epochs, each consisting of 5000 iterations, on an RTX 4090 GPU. The initial learning rate is $4\times10^{-4}$. All the methods are trained with the same setting, except when one fails to converge, in which case the learning rate is reduced. The peak signal-to-noise ratio (PSNR), spectral angle mapping (SAM) \cite{kruse1993spectral}, and SSIM are used to evaluate performance. GFLOPS is calculated with an input image size of $256\times256\times3$.

\subsection{Simulation Results}
\begin{figure*}[t]
  \centering
  \includegraphics[width=\linewidth]{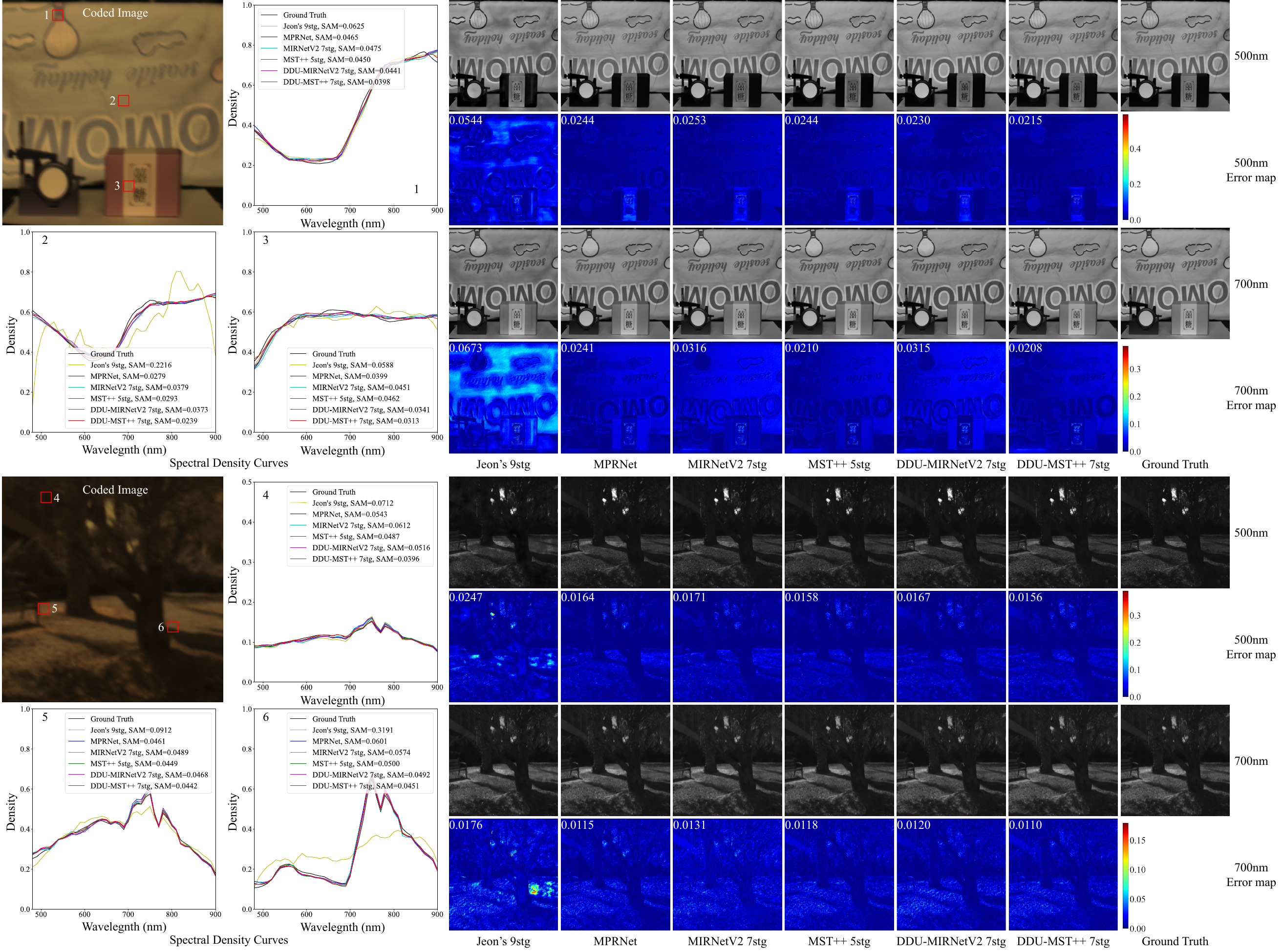}
  \caption{Reconstructed HSI comparisons. The spectral density curves are corresponding to the selected red box of the coded image.  The RMSEs of spectral images compared with GTs are noted in the top-left corner of error maps. Please zoom in. }
  \label{fig: Qual}
\end{figure*}

To validate the effectiveness of our proposed DDU, we evaluate several commonly used networks including HSI estimation methods (HDNet \cite{hu2022hdnet}, MST++ \cite{cai2022mst++}), image restoration models (EDSR \cite{lim2017enhanced}, MPRNet \cite{zamir2021multi}, MIRNetV2 \cite{zamir2022learning}, MambaIR \cite{guo2024mambair}), and deep unfolding methods (ISTANet$^+$ \cite{zhang2018ista}, Jeon's \cite{jeon2019compact}, DPCDUN \cite{song2023dynamic}, CSST \cite{lv2023aperture}). We apply our framework to MST++ and MIRNetV2 separately, named DDU-MST++ and DDU-MIRNetV2. For deep unfolding methods, to indicate the total number of initialization network and iterative blocks, we append the stage number (e.g. 5stg) to each method's name. For MST++ and MIRNetV2, ``stg" denotes the number of SST and RRG blocks, respectively. The results are listed in Table~\ref{tab: quantitative}. To verify whether the gains benefit from more computation cost, we present the results of PSNR vs. FLOPS in Fig.~\ref{fig: Scatter} on mixed test set. 

Our DDU-MST++ with 7 stages shows the best performance, as evidenced by the highest PSNR and SSIM on the ICVL and VNIRSR test sets. It also attains the lowest SAM on the ICVL dataset and performs comparably with the best on the VNIRSR dataset. Specifically, on the VNIRSR test set, it outperforms two SOTA methods, 5-stage MST++ and MPRNet, by 0.38 dB and 0.31 dB in PSNR, respectively.

The previous deep unfolding approaches, ISTANet, Jeon's, and DPCDUN, exhibit poor performance due to two primary reasons. First, their denoiser network architectures are relatively outdated, limiting their reconstruction capabilities. Second, they do not provide an analytical solution to the data fidelity subproblem, leading to suboptimal performance. The latter factor also affects CSST, which replaces data fidelity solver with network layers.

We also observed that MST++ achieves its best performance at 5 stages, while its performance significantly degrades at 7 stages, consistent with the findings in its original work. In contrast, our method prevents such performance degradation, due to the analytical solution we proposed of the data fidelity subproblem. It guides the reconstruction process through stages, leading to a more stable and effective iteration.
After integrating our framework, DDU-MST++ maintains similar Params and FLOPS at the same number of stages while achieving superior performance in all metrics.
Furthermore, our framework also enhances the performance of MIRNetV2, demonstrating that our proposed DDU is compatible with different SOTA architectures.  

The visualization results are shown in Fig.~\ref{fig: Qual}. We show the reconstructed spectral images at 500 nm and 700 nm, along with error maps compared to the ground truths (GTs). To better visualize the spectral reconstruction accuracy, spectral density curves are plotted for selected regions of interest (ROIs). For clarity, the root mean square error (RMSE) and SAM metrics are annotated on the error maps and spectral curves separately.

Obviously, our DDU-MST++ and DDU-MIRNetV2 achieve improved performance compared to the original MST++ and MIRNetV2, respectively. Unlike tasks such as image deblurring, where the primary objective is high spatial resolution, the main challenge in DSSI reconstruction lies in achieving accurate spectral recovery. Reconstruction errors in hyperspectral imaging manifest not only in high-frequency spatial details and textures, but also in the spectral fidelity of low-frequency regions. Accordingly, our DDU-MST++ exhibits the lowest errors in both high-frequency regions (e.g., English letters in the upper image) and low-frequency regions (e.g., ROI2). In low-illumination regions (e.g., ROI4), our framework also exhibits superior performance, demonstrating strong noise robustness. However, ultra-high-frequency structures, such as Chinese characters and grass, remain challenging for all methods.

Overall, our DDU framework improves both visual quality and spectral accuracy, with DDU-MST++ (7-stage) achieving the best performance.

\begin{figure*}[t]
  \centering
  \includegraphics[width=\linewidth]{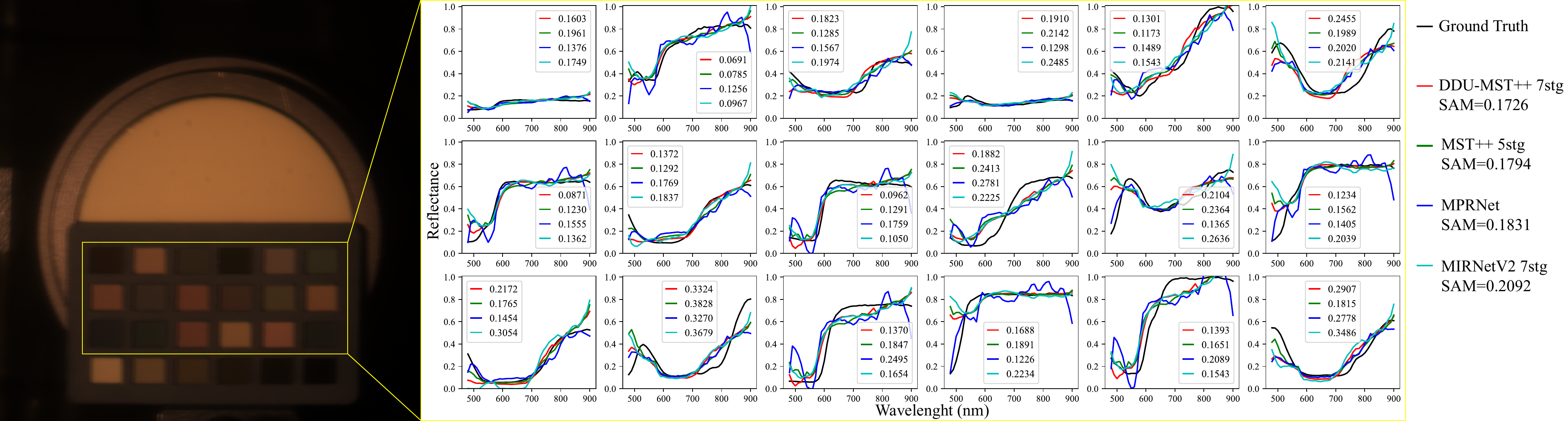}
  \caption{The real-world results of the top three rows of the color checker. The legend in each subplot indicates the SAM values of different methods for each color patch, while the rightmost column displays the average SAM across all selected patches.}
  \label{fig: Real}
\end{figure*}

\begin{figure*}[htpb]
  \includegraphics[width=1.0\linewidth]{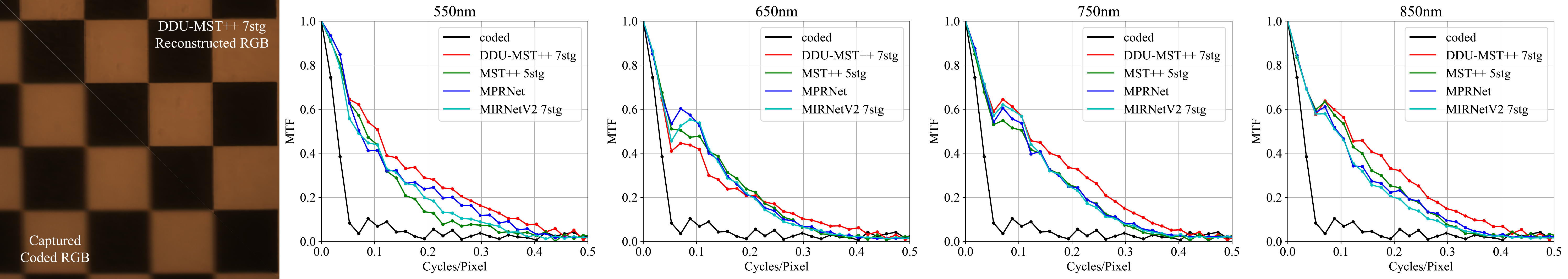}
  \caption{Real captured checkerboard and MTFs at different wavelengths before and after reconstruction.}
  \label{fig: mtf}
\end{figure*}

\subsection{Real-world Results}

In our real-world experiments, the targets were placed near the focal plane of a collimating lens to simulate scenes at virtually infinite distance, illuminated by a halogen lamp. 

A standard whiteboard and a color checker were captured to evaluate the spectral accuracy of reconstruction. To eliminate the influence of the illumination source, the spectra of the color checker patches were normalized by the spectrum extracted from the whiteboard. The reconstructed spectral curves for the top three rows of the color checker are shown in Fig.~\ref{fig: Real}, with ground truth spectra obtained using the GaiaField-V10 HR hyperspectral imager. Among the evaluated methods, our 7-stage DDU-MST++ achieved the lowest average SAM, indicating superior spectral reconstruction. Nonetheless, all methods exhibited some deviations from the ground truth spectra, suggesting a remaining domain gap between simulated and real-world degradation models that warrants further investigation.

To assess the improvement in real-world spatial resolution, we compute the modulation transfer function (MTF) using the slanted edge method, as shown in Fig.~\ref{fig: mtf}. The MTF of the coded measurement is derived from a grayscale image obtained by averaging the RGB channels. As observed, the MTF of the coded input rapidly declines to near zero below 0.1 cycle/pixel, indicating extremely low spatial clarity. All reconstruction methods markedly improve the MTF, with our 7-stage DDU-MST++ yielding the most significant enhancement. Across most wavelengths, the MTF of DDU-MST++ remains non-zero up to approximately 0.5 cycle/pixel, indicating substantially higher spatial fidelity. Apart from a slight drop in the low-frequency region at 650 nm, our method consistently surpasses the others, highlighting the superior capability of the proposed DDU framework in restoring spatial detail from DSSI measurements.

These results confirm that our DDU-MST++ framework effectively recovers both spectral and spatial information from real DSSI measurements.

\section{Ablation Study}

In this section, we systematically analyze the impact of each design choice in our DDU, including the loss function formulation and the detailed architecture of the framework. 

\subsection{Loss Function}

Deep unfolding methods require multi-stage iterative updates, and the effectiveness of gradient propagation may degrade with an increasing number of stages. Therefore, an appropriate loss function is essential to guide network optimization during training. In this section, we evaluate our supervision strategy using a 7-stage DDU-MST++ model.

Table~\ref{tab: loss} compares the performance of the commonly used L1 loss with our adopted L1+SSIM loss. The results show that incorporating the SSIM term improves the SSIM and PSNR while maintaining the SAM. We also investigate the effect of supervising the output at different stages. The results indicate that supervising both $\mathbf{Z}^{(1)}$ and the last output $\mathbf{Z}^{(K)}$ achieves better performance than supervising only $\mathbf{Z} ^{(K)}$ or all intermediate outputs $\{\mathbf{Z}^{(k)}\}_{k=1}^K$. Supervising $\mathbf{Z}^{(1)}$ provides a high-quality initialization that facilitates subsequent iterations, whereas supervising all stages may overly constrain the network and hinder its ability to optimize freely.
\begin{table}[htpb]
\centering
\caption{Ablation study of loss function on DDU-MST++ 5stg.}
\begin{tabular}{ccccc}
\hline
Loss metric    & supervise   & PSNR $\uparrow$ & SAM $\downarrow$ & SSIM $\uparrow$ \\ \hline
L1      & $\{\mathbf{Z}^{(1)},\mathbf{Z}^{(K)}\}$ & 39.36  & 0.0372 & 0.9632          \\
L1+SSIM & $\{\mathbf{Z}^{(1)},\mathbf{Z}^{(K)}\}$ & \textbf{39.40} & \textbf{0.0371} & \textbf{0.9664} \\
L1+SSIM & $\mathbf{Z}^{(K)}$     & 38.88  & 0.0396 & 0.9644\\
L1+SSIM & $\{\mathbf{Z}^{(k)}\}_{k=1}^K$ & 39.37 & \textbf{0.0371} & 0.9658 \\ \hline
\end{tabular}
\label{tab: loss}
\end{table}

\subsection{Framework Structure}
In this section, the experiments involve MST++,  DDU-MST++, and Jeon's model, all uniformly set to 5 stages.

\begin{figure*}[ht]
  \includegraphics[width=1.0\linewidth]{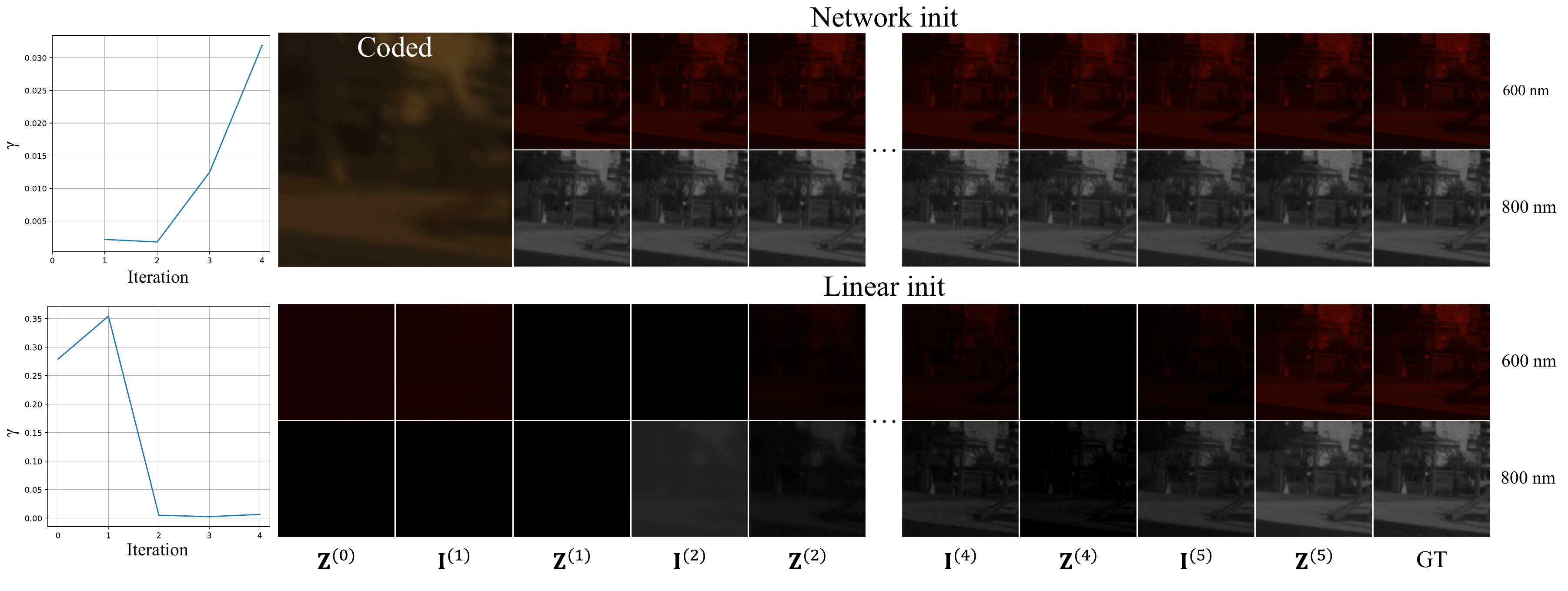}
  \caption{Iteration process with linear and network initialization strategies. The learned parameters $\gamma^{(k)}$ at different stages are shown in the left.}
  \label{fig: iter}
\end{figure*}

\subsubsection{Initialization}
Initialization plays a crucial role in our unfolding method. We evaluate several initialization strategies, as shown in Table~\ref{tab: init}.  For a fair comparison, all baseline strategies are given an additional stage to ensure comparable Params and FLOPS with our default setting, such that baseline methods start from $\mathbf{Z}^{(0)}$. Specifically, ``Rand" and ``Zero" initialization set $\mathbf{Z}^{(0)}$ to random values between $[0,1]$ and zero, respectively. ``Mean" initializes all channels of ``$\mathbf{Z}^{(0)}$" with the mean of $\mathbf{J}$. $\boldsymbol{\Phi}^\mathsf{T}\mathbf{J}$ assigns $\mathbf{Z}^{(0)}=\boldsymbol{\Phi}^\mathsf{T}\mathbf{J}$. ``Linear" employs a learnable linear layer to map $\mathbf{J}$ to $\mathbf{Z}^{(0)}$. Our default initialization $\mathcal{S}_0$ adopts a neural network with a structure similar to $\mathcal{S}$ for this purpose. The results show that among non-learning strategies, ``Mean" yields the best performance. While the ``Linear" mapping improves upon non-learning baselines, it is still inferior to the proposed $\mathcal{S}_0$ network, which achieves a significant performance boost. 

To further investigate the role of initialization, we visualize the iterative behavior of 5 stg DDU-MST++ under linear and network-based initialization in Fig.~\ref{fig: iter}. The learned parameters $\gamma^{(k)}$ at different stages are shown on the left. With network initialization $\mathcal{S}_0$, $\gamma^{(k)}$ gradually increases with iterations, indicating stable convergence. In contrast, under linear initialization, the iteration process exhibits a strong oscillation. $\gamma^{(k)}$ decreases sharply between iterations 1 and 2 and then remains at a relatively low level. Moreover, intermediate outputs exhibit oscillatory behavior, with some values dropping below zero (clipped to [0,255] for visualization). These phenomena reflect instability and poor convergence of linear initialization, which ultimately leads to poor reconstruction quality.
These results highlight the importance of proper initialization in deep unfolding frameworks and demonstrate the effectiveness of directly supervising the initialization network $\mathcal{S}_0$ to ensure stable and high-quality iterative reconstruction.

\begin{table}[htpb]
\centering
\caption{Comparison between different initialization methods.}
\begin{tabular}{ccccc}
\hline
Init. & PSNR $\uparrow$ & SAM $\downarrow$ \\ \hline
Rand         & 38.35 & 0.0424 \\ 
Zero         & 38.40 & 0.0418 \\ 
Mean         & 38.65 & 0.0404 \\ 
$\boldsymbol{\Phi}^\mathsf{T}\mathbf{J}$ & 38.55 & 0.0409 \\ 
Linear      & 38.99 & 0.0398 \\ 
$\mathcal{S}_0$(default) & \textbf{39.31}  & \textbf{0.0372} \\ \hline
\end{tabular}
\label{tab: init}
\end{table}

\subsubsection{ADMM vs. HQS} When we set Lagrangian multipliers $\beta=0 (\widetilde{\boldsymbol{\beta}}=0)$ in our framework, the ADMM unfolding will become the HQS unfolding. $\widetilde{\boldsymbol{\beta}}$ serves as an intermediate variable between $\mathbf{I}$ and $\mathbf{Z}$, which can stabilize the iterative process, as shown in Table~\ref{tab: admm_vs_hqs}. 

\begin{table}[htpb]
\centering
\caption{Comparison between ADMM and HQS unfolding.}
\begin{tabular}{ccccc}
\hline
Method   & PSNR $\uparrow$ & SAM $\downarrow$ \\ \hline
HQS            & 39.23 & 0.0378 \\ 
ADMM(default)  & \textbf{39.31}  & \textbf{0.0372} \\ \hline
\end{tabular}
\label{tab: admm_vs_hqs}
\end{table}

\subsubsection{Linear projection and deep unfolding} In Table~\ref{tab: ablation_of_framework}, we try to add a single linear projection $\mathcal{P}$ with $\gamma=0.5$ after 5-stage MST+, named as ``$\mathcal{P}$+MST++''. It can be observed that our linear projection $\mathcal{P}$, which is the analytical solution Eq.~\ref{eq: solve_U} of the data fidelity subproblem, can directly and slightly improve the reconstruction accuracy. Moreover, to prove the superiority of DDU, we isolate the effect of network architectures. We modify MST++ to ``MST++$^*$", which combines $\mathcal{S}_0$ and $\mathcal{S}$. This is equivalent to removing $\mathcal{P}$ and $\mathcal{M}$ from our DDU-MST++. The results demonstrate that our DDU is effective and efficient, as DDU-MST++ performs the best and introduces only a minimal increase in FLOPS compared to ``MST++$^*$''.

\begin{table}[htpb]
\centering
\caption{Ablation study of linear projection $\mathcal{P}$ and deep unfolding.}
\begin{tabular}{ccccc}
\hline
Method    & Params & GFLOPS   & PSNR $\uparrow$         & SAM $\downarrow$          \\ \hline
MST++     & 11.27M & 151.38 & 39.045         & 0.038811          \\
$\mathcal{P}$+MST++        & 11.27M & 151.44 & 39.048         & 0.038808          \\
MST++$^*$ & 11.17M & 145.04 & 38.704       & 0.040730          \\
DDU-MST++ & 11.17M & 145.26 & \textbf{39.310}  & \textbf{0.037208} \\ \hline
\end{tabular}
\label{tab: ablation_of_framework}
\end{table}

To further validate the superiority of our analytical solution to the data fidelity subproblem, we replace the linear projection step in Jeon's with ours $\mathcal{P}$. For fair comparison, we set $\beta=0 (\widetilde{\boldsymbol{\beta}}=0)$ as Jeon's use HQS unfolding. As shown in Table~\ref{tab: compare_jeon}, our method has significantly better performance with similar FLOPS compared to Jeon's. 

\begin{table}[htpb]
\centering
\caption{Comparison of linear projection step.}
\begin{tabular}{ccccc}
\hline
Method & Params  & GFLOPS           & PSNR $\uparrow$   & SAM $\downarrow$          \\ \hline
Jeon's & 38.73M & \textbf{199.44}    & 30.85          & 0.0992          \\
Ours   & 38.73M  & 199.72 & \textbf{33.61} & \textbf{0.0735} \\ 
\hline
\end{tabular}
\label{tab: compare_jeon}
\end{table}

We also test the speed of our analytical solution of the data fidelity subproblem. For $512\times512\times3$ input, inference times with 7 stages are: MST++ 0.041s, DDU-MST++ 0.054s, and Jeon's-MST++ 0.143s. 
Although MST++ is $24.07\%$ faster than DDU, Jeon's is $164.81\%$ slower, showing that our solver is much faster than original GDM within the DSSI unfolding framework.

\subsubsection{Parameter $\widetilde{\sigma}$ and shared weights}
We evaluate the role of $\widetilde{\sigma}^{(k)}$ and the impact of sharing weights in Table~\ref{tab: shared}. As a result, sharing parameters can effectively reduce the Params with limited affect on performance. When the weights of the network $\mathcal{S}$ are shared across stages, $\widetilde{\sigma}^{(k)}$ can indicate the current stage. When weights are not shared, $\widetilde{\sigma}^{(k)}$ becomes redundant and may hinder optimization.  Thus, we disable $\widetilde{\sigma}$ and do not share the weights in other experiments.

\begin{table}[htpb]
\centering
\caption{Ablation of $\widetilde{\sigma}$ and shared weight.}
\begin{tabular}{cccccc}
\hline
Shared & $\widetilde{\sigma}$ & Params & GFLOPS & PSNR $\uparrow$   & SAM $\downarrow$          \\ \hline
$\times$ & $\times$ & 11.17M & 145.26 & \textbf{39.31} & \textbf{0.0372}          \\
$\times$ & $\checkmark$ & 11.17M & 145.42 & 39.24 & 0.0378   \\
$\checkmark$ & $\times$ & 4.50M  & 145.26 &  38.90 &  0.0401  \\
$\checkmark$   & $\checkmark$ & 4.50M & 145.42 & 39.11 & 0.0388 \\ \hline
\end{tabular}
\label{tab: shared}
\end{table}

\section{Conclusion}
In this paper, we propose an efficient deep unfolding framework, DDU, for DSSI systems. Theoretically, this framework is applicable to any computational spectral imaging system based on PSF encoding that can be represented by Fig.~\ref{fig: Encoding} and Eq.~\ref{eq: degrade}. We derive an analytical solution for the data fidelity subproblem of DSSI and achieve an efficient implementation by leveraging the block diagonal nature of the sensing matrix in the frequency domain. This solver is well-suited for multi-channel to multi-channel deconvolution problems. 

To stabilize the iterative reconstruction process, we use a neural network to initialize the input. Supervision is applied to both the initialization and the final output to promote better optimization. Furthermore, we conduct an in-depth analysis of the loss function and the architecture of the DDU framework to enhance overall performance. 

Overall, our DDU framework is highly compatible with SOTA networks and achieves superior results while maintaining similar Params and FLOPS. In real-world experiments using our prototype camera, our DDU-MST++ with 7 stages achieves the best spectral and spatial reconstruction results. We believe that this framework provides a solid foundation for future deep unfolding methods in DSSI.

\bibliographystyle{IEEEtran}
\bibliography{main}

\end{document}